\documentclass[12pt]{article}

\usepackage{graphicx} 
\usepackage{jheppub}
\usepackage{amsfonts,amsmath} 
\usepackage{mathrsfs} 
\usepackage{setspace}
\usepackage[utf8]{inputenc} 
\usepackage{ytableau}
\usepackage{comment} 

\makeatletter
\def\@fpheader{\relax}
\makeatother

\newcommand{\be}{\begin{equation}} \newcommand{\ee}{\end{equation}}

\newcommand{\thalf}{{\tfrac{1}{2}}}
\DeclareMathOperator{\Det}{Det}

\DeclareMathOperator{\diag}{diag}
\DeclareMathOperator{\Tr}{Tr}

\DeclareMathAlphabet{\mathbbold}{U}{bbold}{m}{n} 

\title{Kerr Scattering Coefficients via Isomonodromy}

\author[a]{Bruno Carneiro da Cunha}\emailAdd{bcunha@df.ufpe.br}
\author[b]{and Fábio Novaes}\emailAdd{fabio.nsantos@gmail.com}

\affiliation[a]{Departamento de Física, Universidade Federal de
  Pernambuco, 50670-901, Recife, Pernambuco, Brazil}
\affiliation[b]{International Institute of Physics, Federal University
  of Rio Grande do Norte, Av. Odilon Gomes de Lima 1722, Capim Macio,
  Natal-RN 59078-400, Brazil} 

\abstract{We study the scattering of a massless scalar field
  in a generic Kerr background. Using a particular gauge choice based
  on the current conservation of the radial equation, we give a
  generic formula for the scattering coefficient in terms of the
  composite monodromy parameter $\sigma$ between the inner and the
  outer horizons. Using the isomonodromy flow, we calculate $\sigma$
  exactly in terms of the Painlevé V $\tau$-function. We also show
  that the eigenvalue problem for the angular equation (spheroidal
  harmonics) can be calculated using the same techniques. We use
  recent developments relating the Painlevé V $\tau$-function to
  Liouville irregular conformal blocks to claim that this scattering
  problem is solved in the combinatorial sense, with known expressions
  for the $\tau$-function near the critical points.} 


\preprint{\today}

\begin{document}

\maketitle

\section{Introduction}

Black hole scattering theory is an important problem from the general
relativity and string theory perspective. It serves as a way of
testing linear stability \cite{Dafermos:2008en}, superradiance
\cite{Press:1972zz}, black hole entropy \cite{York:1983zb} and
relaxation times in AdS-CFT \cite{Horowitz:1999jd}. For the rich story
behind it see \cite{Kokkotas:1999bd,Berti:2009kk}. Scattering is also
important to study astrophysical problems related to the detection of
gravitational waves and a possible mechanism for short gamma ray
bursts, for example. See the 
introduction of \cite{Dolan:2008kf}, references therein and
\cite{Fiziev:2009ty,Kumar:2014upa} for discussions. It is also a daunting problem
from the mathematical side, involving unknown properties of new
special functions \cite{Borissov:2009bj}. Most of the previous study
in four dimensions on the subject relied on numerical analysis, either
by direct numerical integration or by expansion of the solution in
terms of known special functions
\cite{Leaver:1985ax,Mano:1996vt,Mano:1996gn}. Also, decoupling limits
like the near-extremal cases were taken directly on the metric (see
\cite{compere2012kerr} for applications in Kerr-CFT).

In this letter we build on previous analytical work, a
reverse Hilbert-Riemann problem \cite{Castro2013b}, based on
isomonodromy equations \cite{Novaes:2014lha} to relate the scattering
coefficients for a massless scalar field in four-dimensional Kerr
background in terms of the Painlevé V transcendent $\tau$-function
\cite{Gamayun:2013auu} and show that the problem can be solved in the
combinatorial sense using the expasions of the $\tau$-function near
the critical points. We also show that the angular equation can be
solved implicitly using the same techniques, in a new way of
extracting eigenvalues for the spheroidal wave functions
\cite{Flammer2014,Berti:2005gp}. The results shown here can be understood as the
natural conclusion to the application of the monodromy problem to
solve scattering problems as applied to black holes
\cite{Neitzke:2003mz,Andersson:2003fh,Motl:2003cd,Keshet:2007be}. The
result follows from a careful consideration of the ``linearization'' of the Painlevé V
equation in the sense given by Miwa, Jimbo {\it et al.} in
\cite{Jimbo1981b,Jimbo:1981-2,Jimbo:1981-3}, and of the time-reversal
symmetry. These two points allows us to place boundary conditions such
that the two linearly independent solutions of the radial equation
correspond to the two solutions involved in the connection problem of
the Painlevé V equation \cite{Jimbo:1982}. Recent developments on the
relation between Painlevé transcendents and the AGT conjecture
\cite{Alday:2009aq,Alba:2010qc} show the striking connection between
black holes and black hole scattering to Conformal Field Theory
\cite{bredberg2010black}.

This letter is organized as follows. In Section \ref{sec:kg-kerr} we
treat the radial equation for the massless Klein-Gordon field in a
generic four-dimensional Kerr background. We write the equation in a
canonical confluent Heun type and discuss the time-reversal
symmetry. In Section \ref{sec:fuchsian} we discuss how the scattering
problem for the radial equation is connected to the connection problem
of the confluent Heun equation. We describe the (extended) monodromy
data of the confluent Heun equation, including the Stokes parameters,
and use the time-reversal symmetry to parametrize the scattering
coefficient in terms of the monodromy parameters. In Section
\ref{sec:isomonodromy} we discuss how to extract these parameters from
the Painlevé V $\tau$-function, using the isomonodromy flow. In
Section \ref{sec:angulareq}, we outline a procedure to extract the
eigenvalues for the angular equation using the same
structure. Finally, in Section \ref{sec:painleve5} we list the results of
\cite{Gamayun:2013auu} which give a combinatorial expansion for the
Painlevé V $\tau$-function near $t_0=0$, which would correspond to our
extremal limit. We close with a short discussion and future directions
in Section \ref{sec:discussion}. 

\section{Klein-Gordon in Kerr}
\label{sec:kg-kerr}

It is known that the massless Klein-Gordon equation in a Kerr
background is separable \cite{Carter:1968kx}, {\it i. e.}, the solution can
be written as $\Phi(x^\mu)=e^{-i\omega t}e^{im\phi}
R(r)S(\theta)$.
The angular part $S(\theta)$ is solved in terms of oblate scalar
spheroidal harmonics, whose eigenvalues (the separation constant
$C_{\ell,m}$) can be approximated for different asymptotic regimes,
including low and high frequency limits
\cite{Flammer2014,NISTSpheroidal}, and will be the object of study in
Section \ref{sec:angulareq}.  We will focus for now on the radial part
\begin{gather}
  \label{eq:kgAdS-4d}
  \partial_r(Q(r)\partial_r R(r)) + \left(-C_{\ell,m} +
    \frac{W_{r}^2}{Q(r)}\right) R(r) = 0\;,
\end{gather}
where $Q(r)$ and $W_r^2$ are given by
\begin{equation}
\begin{aligned}
  \label{eq:kerr-polynomial}
  Q(r) &= r^2 - 2Mr + a^2=(r-r_+)(r-r_-), \\[5pt]
   W_{r} &= \omega(r^2 + a^2)-am.
\end{aligned}
\end{equation}
The form which will be useful for the analysis has
\begin{equation}
\label{eq:1}
z=2i\omega(r-r_-),\quad\text{ and }\quad
y(z)=(r-r_-)^{\theta_0/2}(r-r_+)^{\theta_{t_{0}}/2}R(r),
\end{equation}
with the parameter
\begin{equation}
t_0=2i\omega(r_+-r_-).
\end{equation}
The radial equation is written in this parametrization as
\begin{equation}
\begin{gathered} 
\frac{d^2y}{dz^2}+p(z)\frac{dy}{dz}+q(z)y(z)=0,\\[10pt]
p(z)=\frac{1-\theta_{t_0}}{z-t_0}+\frac{1-\theta_0}{z},\quad
q(z)=-\frac{1}{4}+\frac{c_0}{z}+\frac{c_{t_0}}{z-t_0},
\label{eq:heunconfluent}
\end{gathered}
\end{equation}
which is known as the confluent Heun equation, with parameters given
by
\begin{equation}
\begin{gathered}
\theta_0 =2i\frac{\omega(r_{-}^2+a^2)-am}{r_--r_+}
=-\frac{i}{2\pi}\frac{\omega-\Omega m}{T_-},\\[5pt]
\theta_{t_0} =2i\frac{\omega(r_{+}^2+a^2)-am}{r_+-r_-}=
\frac{i}{2\pi}\frac{\omega-\Omega m}{T_+}, \\[5pt]
t_0c_0 =C_{\ell,m}+\frac{1}{2}((1-t_0)\theta_0+\theta_{t_0}), \quad\quad
t_0c_{t_0}=-C_{\ell,m}-\frac{1}{2}(\theta_0+(1+t_0)\theta_{t_0}),
\label{eq:heunconfluentparameters}
\end{gathered}
\end{equation}
where we defined the temperatures $T_\pm$ and angular velocities
$\Omega_\pm$ for the outer and inner horizons. For future reference, we
also define 
\begin{equation}
\quad\quad
%\theta_\infty=1-2i\omega(r_++r_-)=1-\theta_0-\theta_{t_0}.
\theta_\infty = 1+2(c_0+c_{t_0})=1-4iM\omega
=1-\theta_0-\theta_{t_0}.
\label{eq:inftydata}
\end{equation}
Also, eq.~\eqref{eq:kgAdS-4d} has real coefficients, so if $R(r)$ is a
solution, so is $R(r)^*$. We then define the radiation flux as
\begin{equation}
  \label{eq:current}
  j=-i\, Q(r)[R(r)^*\partial_r R(r)-R(r)\partial_r R(r)^*].
\end{equation}
% , written in
% terms of $y(z)$ as
% %\begin{equation}
% \begin{multline}
% j=i\frac{z(z-t_0)}{4\omega^{2}}e^{-i\pi(\theta_{0}+\theta_{t_{0}})/2}\left[P(z)y^*(z)\frac{d}{dz}(P(z)^{-1}y(z))
%  \right. \\-
% \left. P(z)^{-1}y(z)\frac{d}{dz}(P(z)y^*(z))\right],
% \label{eq:currentdensity}
% \end{multline}
% %\end{equation}
% where $P(z)=z^{\theta_0/2}(z-t_0)^{\theta_{t_0}/2}$.
One should bear
in mind that real $r$ means $z^*=-z$, because of \eqref{eq:1}.

\section{The Fuchsian System}
\label{sec:fuchsian}

Writing the solutions for \eqref{eq:heunconfluent} in terms of
Frobenius expansions at $z=0$ or $z=t_0$ is straightforward. On the
other hand, extracting the scattering coefficients from the solutions
is not straightforward in the sense that not enough information is
known about the function defined by the ODE. To wit, we are interested
in the so-called {\it connection problem}: how a solution of 
\eqref{eq:heunconfluent} with known behavior at one singular point,
say $y_{t_0}(z)$ at $z=t_0$, is related to the pair of solutions with
known behavior at another singular point, say, $y_\infty^\pm(z)$ at
$z=\infty$. So, for example, in
\begin{equation}
y_{t_0}(z)=C y_\infty^+(z)+D y_\infty^-(z),
\label{eq:connectiondata}
\end{equation}
$C$ and $D$ are the connection coefficients.  If $y_{t_0}(z)$
represents a purely ingoing wave at the outer horizon $z=t_0$ and
$y_\infty^\pm(z)$ ingoing and outgoing waves at infinity, then the
connection coefficients are indeed the scattering coefficients. As we
will see below, solving the connection problem allows us to solve the
scattering problem.

In \cite{Novaes:2014lha}, the authors proposed a method based on the
Schlesinger equations to extract the scattering coefficients. It
relies on viewing the ODE \eqref{eq:heunconfluent} as the equation
satisfied by the first row of the $2\times 2$ matricial Fuchsian
system
\begin{equation}
\frac{d}{dz}\Phi(z)=A(z)\Phi(z),
\label{eq:schlesinger}
\end{equation}
and to extract the connection coefficients from the monodromy data of
the Fuchsian system. The monodromy data for the system at hand is
defined as follows: suppose $A(z)$ is meromorphic and let $\{z_i\}$
be the isolated singular points of the system
\eqref{eq:schlesinger}. The point at $z=\infty$ is studied by the
substitution $u=1/z$. The solution $\Phi(z)$ will also cease to be
analytic at these points and behave singularly there. Since
$\tilde{\Phi}(z)=\Phi(e^{2\pi i}(z-z_i)+z_i)$ is also a solution of
\eqref{eq:schlesinger}, then we must have
$\tilde{\Phi}(z)=\Phi(z)M_i$, with $M_i$ a constant matrix associated
with a circuit around $z_i$ -- the monodromy.

The form of $M_i$ depends on the behavior of $A(z)$ near the singular
point $z_i$. If $A(z)$ diverges like $(z-z_i)^{-1}$, then the singular
point is called regular and $M_i$ is conjugated to
$\diag(e^{2\pi i\alpha^+_i},e^{2\pi i\alpha^-_i})$, where
$\alpha^\pm_i$ are the solutions of the indicial equation near $z_i$.
This form of $M_i$ stems from the leading behavior
$(z-z_i)^{\alpha^\pm_i}$ of the solution near $z_i$, a branch point
for $\Phi(z)$. We will for now on assume that the differences
$\alpha^+_i-\alpha^-_i$ are not integers for the radial equation. If,
however, $A(z)$ diverges like $(z-z_i)^{-1-r_i}$, where the integer
$r_i>0$ is the Poincaré rank, the solution will have an essential
singularity and will experience the {\it Stokes phenomenon}: the
leading term in the Frobenius expansion displays an exponential
behavior which diverges or converges depending on which direction the
limit $z\rightarrow z_i$ is taken in the complex plane. In order to
describe it, we will specialize to our case. In equation
\eqref{eq:heunconfluent}, the matricial system has $A(z)$ with the
structure
\begin{equation}
A(z)=\frac{1}{2}\sigma_3+\frac{A_0}{z}+\frac{A_t}{z-t},
\label{eq:heunschlesinger}
\end{equation}
where there are two regular singular points, at $z=0$ and $z=t$, and
an irregular point at $z=\infty$ with Poincaré rank 1. We also have
the usual Pauli matrix $\sigma_3$. The elements of $A_0$ and $A_t$ in
\eqref{eq:heunschlesinger} are independent of $z$ and are obtained
from the ODE \eqref{eq:heunconfluent}. This identification has extra
freedom because the Fuchsian system has more free parameters than the
ODE. We thus pick the \emph{gauge choice}
\begin{equation}
  \Tr A_i=\theta_i,\quad \det A_i= 0,\quad \Tr
  \sigma_3(A_0+A_t)=-\theta_\infty, 
\label{eq:conservedquantities}
\end{equation}
which, along with the ODE parameters
\eqref{eq:heunconfluentparameters}, fixes the elements of $A_i$ up to
conjugation by $e^{u\sigma_3}$. The
epithet \emph{gauge choice} will become clear in the next Section.

In order to describe the system at the irregular singular point $z=\infty$, one
defines the sectors \cite{Jimbo:1982,Andreev2000}
\begin{equation}
{\cal S}_j=\left\{z\in\mathbb{C}\, |\, (2j-5)\frac{\pi}{2} < \arg z <
(2j-1)\frac{\pi}{2}\right\}, 
\end{equation}
for $j\in \mathbb{Z}$. On each ${\cal S}_j$, one defines the asymptotic
behavior for $\Psi(z)=z^{-\thalf(\theta_0+\theta_t)}\Phi(z)$ based on
the gauge choice \eqref{eq:conservedquantities} for $A(z)$, 
\begin{equation}
\Psi(z)\big|_j=(\mathbbold{1}+{\cal O}(z^{-1}))\exp(\thalf
z\sigma_3)z^{-\thalf \theta_\infty \sigma_3}.
\label{eq:boundaryinfty}
\end{equation}
These canonical solutions are connected by the Stokes matrices
$\Psi_{j+1}(z)=\Psi_j(z)S_j$ and, given that $\Psi_{j+2}(e^{2\pi
  i}z)=\Psi_j(z)e^{-\pi i \theta_\infty\sigma_3}$, 
we have
$S_{j+2}=e^{\pi i \theta_\infty\sigma_3}S_j e^{-\pi i
  \theta_\infty\sigma_3}$,
and the Stokes matrices are all determined by any consecutive 2, which
have the following structure
\begin{equation}
S_{2j}=
\begin{pmatrix}
1 & s_{2j} \\
0 & 1
\end{pmatrix},\quad
S_{2j+1}=
\begin{pmatrix}
1 & 0 \\
s_{2j+1} & 1
\end{pmatrix},
\end{equation}
and the $s_{2j}$ and $s_{2j+1}$ are called {\it Stokes
  parameters}. Local solutions of the Fuchsian system at $z=\infty$
are defined in the Riemann surface of the logarithm, and these
parameters describe how the solution \eqref{eq:boundaryinfty} changes
as one changes the leaf of the surface. The (extended) monodromy data for
\eqref{eq:heunconfluent} is completely determined by the monodromy
matrices $M_0$ and $M_t$, at the regular singular points $0$ and $t$,
respectively, and the Stokes matrices $S_0$ and $S_1$, which determine
the (formal) monodromy matrix at $\infty$ in sector ${\cal S}_j$ by
\begin{equation}
M_\infty\big|_{{\cal S}_j}=S_jS_{j+1}e^{\pi i \theta_\infty\sigma_3}.
\end{equation}
Knowledge of the $\theta_i$ and of two consecutive Stokes parameters, say,
$s_1$ and $s_2$ is sufficient data for solving the connection
problem. We define the connection matrices $E_i$ by
\begin{equation}
M_i=E_i^{-1}
\begin{pmatrix}
e^{i\pi\theta_i} & 0 \\
0 & e^{-i\pi\theta_i}
\end{pmatrix}
E_i,\quad\quad i=0,1,
\label{eq:connectionmatrices}
\end{equation}
and those are given in terms of $\theta_i$, $\sigma$, $s_1$ and $s_2$
in \cite{Jimbo:1982,Andreev2000}. Note that, because
$M_\infty M_{t_{0}} M_0= \mathbbold{1}$, we have that the combined
monodromy parameter $2\cos\pi\sigma=\Tr M_0M_t$ satisifies
\begin{equation}
2\cos\pi\sigma = 2\cos\pi\theta_\infty +
e^{i\pi\theta_\infty}s_1s_2,
\label{eq:sigmastokes}
\end{equation} 
and that
$ M_{\infty} = E_\infty^{-1} e^{i\pi\sigma \sigma_{3}}E_\infty$.  The
connection matrices $E_i$ can be seen -- when properly normalized by
$\det E_i=1$ -- as the matrices that implement the change to the
``natural basis'' of solutions of the ODE at $z_i$ where the monodromy
$M_i$ is diagonal. Explicit formulas for $E_i$ in terms of monodromy
data are given in \cite{Andreev2000}. We use the remaining gauge
symmetry $A(z)\rightarrow e^{u\sigma_3}A(z) e^{-u\sigma_3}$ so that both
entries in the first row of $\Phi(z)$ have equal but opposite current
density normalization \eqref{eq:current}, defined as $j^\pm$. So we
have the entries $[\Phi(z)]_{11} = u^{+}_\infty \propto R^{+}_\infty$
and $[\Phi(z)]_{12} = u^{-}_\infty \propto R^{-}_\infty$ being the
normalized incoming and outgoing waves near infinity. Because the
$j^\pm$ are conserved, this choice also means that the entries will be
normalized near the outer horizon $t_0$, up to a phase. Therefore, the
entries of the connection matrix 
\begin{equation}
  \label{eq:connectionmatrix}
  E_{\infty t_{0}} = E_{\infty}E_{t_{0}}^{-1}
\end{equation}
will give the
explicit scattering coefficients, allowing us to write ${\cal T}$ and
${\cal R}$ such that
\begin{equation}
u^{-}_{t}=\frac{1}{{\cal T}}u^{+}_\infty+\frac{{\cal R}}{{\cal T}}u^{-}_\infty,
\label{eq:scatteringdata}
\end{equation}
and by construction the time-reversed wave is equal, up to a
phase, to the entry $[\Phi(z)]_{12}$ near $t_0$. We notice that
$\Det E_{\infty t_0}=1$ implies $|{\cal R}|^2=1-|{\cal T}|^2$, which
is the condition of flux conservation. Now, by substitution of the
diagonalized forms of the monodromy matrices in
$\Tr M_\infty M_t=\Tr M_0^{-1}$, and using \eqref{eq:scatteringdata},
we obtain the formula
\begin{equation}
|{\cal T}|^2=\left|\frac{\sin \pi\sigma \sin \pi \theta_{t_{0}}}
{\sin\frac{\pi}{2}(\theta_{0}+\sigma-\theta_{t_{0}})\sin
  \frac{\pi}{2}(\theta_{0}-\sigma+\theta_{t_{0}})}\right|, 
%{\cos\pi(\sigma-\theta_0)-\cos\pi \theta_{t_{0}}},
% |{\cal T}|^2=\left|\frac{2\sin \pi\theta \sin \pi \theta_{t_0}}{
%   C^{-1/2}\cos\pi\theta_0+\cos\pi (\theta-\theta_{t_0})}\right|,
\label{eq:transmissioncoeff}
\end{equation}
%where $C=1+s_1s_2$ and $\theta=\theta_{\infty}+(i\pi)^{-1}\log C$. 
which gives the coefficient in terms of the composite monodromy
parameter $\sigma$. In order to compute it, we will need the structure
of isomonodromy deformations, outlined in the next section. After
embedding this system in the isomonodromy formulation, we will be able
to give explicit coordinate dependence of the wave functions.

\section{The Isomonodromy Method}
\label{sec:isomonodromy}

The idea of considering the Fuchsian system \eqref{eq:schlesinger}
instead of the scalar ODE came in the context of the Riemann-Hilbert problem,
since a simple counting argument \cite{Iwasaki:1991} shows that the
generic scalar ODE does not have enough parameters to realize all
possible monodromy matrices. The Fuchsian system, however, has too
many parameters, so 
there is a family of $A(z)$ in \eqref{eq:schlesinger} with the same
monodromy data. The description of this family can be
understood physically: suppose we are given an invertible solution
$\Phi(z)$ of \eqref{eq:schlesinger} with some singular points. Then we
can understand $A(z)=[\frac{d}{dz}\Phi(z)]\Phi(z)^{-1}$ as a ``gauge
potential'' describing non-abelian charges, dipoles or multipoles at
each singular point. The physical configuration described by the gauge
potential is invariant under generic gauge transformations
$A(z)\rightarrow U(z)A(z)U(z)^{-1}+[\frac{d}{dz}U(z)]U(z)^{-1}$,
meaning that the monodromy data (non-abelian holonomies) are
also invariant under these transformations. 

In the case at hand \eqref{eq:heunconfluent}, with $A(z)$ of the
form \eqref{eq:heunschlesinger}, we supply a $t$-component to this
gauge potential,
\begin{equation}
B(z)=-\frac{A_t}{z-t},
\end{equation}
in such a way that the ``field strength''
$\partial_tA-\partial_zB+[A,B]$ is zero if the {\it Schlesinger
  equations} hold
\begin{equation}
\frac{\partial A_0}{\partial t}=\frac{1}{t}[A_t,A_0],\quad
\frac{\partial A_t}{\partial t}=-\frac{1}{t}[A_t,A_0]-\frac{1}{2}[A_t,\sigma_3].
\label{eq:schlesingerheun}
\end{equation}
The monodromy data of the Fuchsian system \eqref{eq:schlesinger} will
be independent of $t$ if $A_0(t)$ and $A_t(t)$ in
\eqref{eq:heunschlesinger} satisfy the equations above
\eqref{eq:schlesingerheun}. For the relationship between the
Schlesinger equations and the theory of flat holomorphic connections
we refer to \cite{Jimbo1981b,Jimbo:1981-2,Jimbo:1981-3,Iwasaki:1991}.

The system \eqref{eq:schlesinger} has some direct conserved
charges. The gauge choice \eqref{eq:conservedquantities} is justified
now because these are the constant of motions under Schlesinger
evolution. With these constraints, the equation satisfied by any
element of the first row of $\Phi(z)$ in \eqref{eq:schlesinger}, with
$A(z)$ given by \eqref{eq:heunschlesinger}, is of the form
\begin{equation}
\begin{gathered}
\frac{d^2y}{dz^2}+p(z)\frac{dy}{dz}+q(z)y=0, \\[5pt]
p(z)=\frac{1-\theta_0}{z}+\frac{1-\theta_t}{z-t}-\frac{1}{z-\lambda},
\quad\quad
q(z)=-\frac{1}{4}+\frac{C_0}{z}+\frac{C_t}{z-t}+\frac{\mu}{z-\lambda},
\label{eq:edofromschlesinger}
\end{gathered}
\end{equation}
%\begin{equation}
%\begin{gathered}
% C_0 =
% \frac{\lambda(\lambda-t)}{t}\left[\mu^2-
%\left(\frac{\theta_0-1}{\lambda} +\frac{\theta_t}{\lambda-t}\right)
%+\frac{\theta_\infty-1}{2(\lambda-t)}-\frac{1}{4}\right] \\
%C_1=-\frac{\lambda(\lambda-t)}{t}\left[\mu^2-
%\left(\frac{\theta_0}{\lambda} +\frac{\theta_t-1}{\lambda-t}\right)
%-\frac{\theta_\infty-1}{2(\lambda-t)}-\frac{1}{4}\right]
%+\frac{1}{2}(\theta_\infty-1) 
%\end{gathered}
%\end{equation}
where $C_0$, $C_t$, $\lambda$ and $\mu$ are functions of the entries of
$A(z)$. Compared to \eqref{eq:heunconfluent}, the system above has an
extra singularity at $z=\lambda$. However, this is an
apparent one --  its monodromy is trivial, as the indicial equation has
exponents $0$ and $2$ and there is no logarithm behavior due to the
relation between $\mu$, $\lambda$ and $t$,
\begin{equation}
\mu^2-\left[\frac{\theta_0-1}{\lambda}+\frac{\theta_t-1}{\lambda
    -t} \right]\mu+\frac{C_0}{\lambda}+\frac{C_t}{\lambda-t}=\frac{1}{4}.
\label{eq:algebraicconstraint}
\end{equation}
The system \eqref{eq:schlesingerheun} is better known in another
clothing, as
\begin{equation}
y(t)=\frac{[A_0(t)]_{11}[A_t(t)]_{12}}{[A_t(t)]_{11}[A_0(t)]_{12}}= 
\frac{\theta_0+\theta_t-\theta_\infty-(2\mu-1)(\lambda-t)}{
  \theta_0+\theta_t-\theta_\infty-(2\mu-1)\lambda}  
\end{equation}
can be checked to satisfy the Painlevé V equation. These are part of
the Painlevé transcendents family of differential equations:
non-linear second order differential equations whose solutions do not
possess movable branch points \cite{Iwasaki:1991}. These define new
special functions, with applications in integrable systems, random
matrix theory and conformal field theory -- see
\cite{Gamayun:2012ma,Gamayun:2013auu} for references in those
applications. In \cite{Jimbo:1982}, asymptotic expressions for the
Painlevé V system were given in the guise of the $\tau$-function
\begin{equation}
\begin{aligned}
\frac{d}{dt}\log\tau(t,\{\theta_i\},s_1,s_2) & =
-\frac{1}{2}\Tr\sigma_3A_t -\frac{1}{t}\Tr A_0A_t \\
%& = C_t+\frac{\mu\lambda}{t}-\frac{\lambda-t}{2t}-\frac{\theta_0\theta_t}{t}, 
& =-\frac{\lambda(\lambda-t)}{t}\left[
\mu^2-\left(\frac{\theta_0}{\lambda}+\frac{\theta_t}{\lambda-t}\right)\mu
+\frac{\theta_\infty}{2\lambda}-\frac{1}{4}\right]-\frac{\theta_0\theta_t}{t},
\label{eq:ourtaufunction}
\end{aligned}
\end{equation}
in terms of the monodromy data. The $\tau$-function is the most
natural isomonodromy invariant that can be defined, with a clear
interpretation as a generating functional in quantum field theory
applications \cite{jimbo1980density,Segal1985,Mason:2000aa}.

As the problem now stands, knowledge of the $\tau$ function solves the
system completely. We pick suitable initial conditions for $\mu$ and
$\lambda$ in \eqref{eq:edofromschlesinger}, and from those recover the
$A_0$ and $A_t$, which then will set the initial conditions for the
$\tau$ function. By choosing the initial conditions
\begin{equation}
\theta_t=\theta_{t_0}-1,\quad\lambda(t_0)=t_0,\quad\mu(t_0)=
\frac{c_{t_0}}{\theta_{t_0}-1}, 
\label{eq:initialcondition}
\end{equation}
based on the parameters of the initial Heun equation
\eqref{eq:heunconfluentparameters} we have 
\begin{equation}
\begin{aligned}
\left.t\frac{d}{dt}\log\tau(t;\{\theta_i\},s_i)\right|_{t=t_0}
&=t_0c_{t_0}-\theta_0(\theta_{t_0}-1) ,\\[5pt]
\left.t\frac{d}{dt}\left(t\frac{d}{dt}\log\tau(t;\{\theta_i\},s_i)\right)\right|_{t=t_0}
&=t_0\frac{\theta_{t_0}-1}{2},
\label{eq:initialdata}
\end{aligned}
\end{equation}
which can be formally inverted to yield the Stokes parameters $s_1$
and $s_2$, and hence $\sigma$. We note that our definition
\eqref{eq:ourtaufunction} differs slightly from the one in
\cite{Jimbo:1982} by
\begin{equation}
\tau_{\rm
  ours}(t)=t^{((\theta_0-\theta_t)^2-2\theta_\infty^2)/4}[\tau_{\rm
  Jimbo}(t)]^{-1}.
\label{eq:jimbodefs}
\end{equation}
One also notes that \eqref{eq:initialdata} allows us to interpret the
$\tau$-function as the generating function that implements the
canonical transformation between the non-trivial monodromy data
$s_1,s_2$ and the canonically conjugated parameters of the ODE
$t_0,c_{t_0}$. 

The structure outlined here gives another interpretation for the
$\tau$-function. One of the most fundamental results from Miwa, Jimbo
{\it et al.} on the Schlesinger system
\cite{Jimbo1981b,Jimbo:1981-2,Jimbo:1981-3} -- see also
\cite{Iwasaki:1991} -- was the Hamiltonian structure of the
isomonodromy flow. This structure stems from the symplectic structure
of  flat holomorphic connections (see \cite{Novaes:2014lha} for a
description), but can be intuitively understood from the algebraic
constraint \eqref{eq:algebraicconstraint}. One first notes that $C_0$
and $C_t$ are not independent, they are related to the asymptotic
behavior at $z=\infty$, via $\theta_\infty$,
\begin{equation}
C_0+C_t=-\mu+\frac{\theta_\infty-1}{2}.
\end{equation}
Varying $t$ will now tie the
variation of $\lambda$ and $\mu$ in a symplectic system, in which
$C_0$ can be thought of as the Hamiltonian function. Now, by
``solving'' the Hamiltonian system one means giving a canonical
transformation from the coordinates $C_0,t_0$ to coordinates where the
isomonodronic flow is trivial. But the latter is parametrized by the
monodromy data, which by construction is invariant under the flow. The
$\tau$-function is then the generating function that implements this
canonical transformation, as evidenced by \eqref{eq:initialdata}. In
\cite{Nekrasov:2011bc} a set of Darboux coordinates were constructed
for the Painlevé VI case. It would be interesting to give a parallel
of that construction to the case at hand.

\subsection{Asymptotic wavefunctions and normalization}

Now, with the embedding of the radial equation \eqref{eq:kgAdS-4d} in
the matricial system \label{eq:schlesinger} via
\eqref{eq:initialcondition}, we can work out the specific form of the
wave functions in terms of the coordinate $r$. The form is not
necessary for our results, but will help comparing with previous work
\cite{Mano:1996vt,Mano:1996gn}. The normalization of
the elementary matrix $\Phi(z)$ is given by 
\begin{equation}
\det \Phi(z)= z^{\theta_0}(z-t_0)^{\theta_{t_0}-1},
\end{equation}
with boundary conditions at $z=\infty$ given by \eqref{eq:boundaryinfty}. It is
straightforward to verify that the Wronskian between the two entries
of the first row is given by:
\begin{equation}
W(z)=[\Phi'(z)]_{11}[\Phi(z)]_{12}-[\Phi'(z)]_{12}[\Phi(z)]_{11}=[A(z)]_{12}\det\Phi.
\end{equation}
Now, using the form of $[A(z)]_{12}$ compatible with
\eqref{eq:edofromschlesinger}, with conditions \eqref{eq:initialcondition}:
\begin{equation}
[A(z)]_{12}=\frac{k(z-\lambda)}{z(z-t)}=\frac{k}{z}
\end{equation}
we arrive at:
\begin{equation}
W(z)=k z^{\theta_0-1}(z-t_0)^{\theta_{t_0}-1}.
\end{equation}
We remind the reader that the value $k$ is ``gauge
dependent'', and proceed to
normalize the asymptotic from the entries of $\Phi(z)$. From
\eqref{eq:boundaryinfty}, we have: 
\begin{equation}
\begin{gathered}
\mbox{} [ \Phi_{\infty} (z) ]_{11}= z^{\thalf(\theta_0+\theta_{t_0}-1-\theta_\infty)}e^{\thalf
  z}(1+{\cal O}(z^{-1})), \\
[\Phi_{\infty}(z)]_{12}=k z^{\thalf(\theta_0+\theta_{t_0}-1+\theta_\infty)-1}e^{-\thalf
  z}(1+{\cal O}(z^{-1})),
\end{gathered}
\end{equation}
where the limit is taken through the
direction $\arg z = \thalf\pi$ ($z\rightarrow +i\infty$). Following
\cite{Jimbo:1982} we define the Frobenius basis at $z=t_0$ to be:
\begin{equation}
\Phi_{t_0}(z)=G_{t_0}
\begin{pmatrix}
1 & 0 \\
0 & (z-t_0)^{\theta_{t_0}-1}
\end{pmatrix}
(\mathbbold{1}+{\cal O}(z-t_0)),
\end{equation}
where $G_{t_0}$ is a matrix which diagonalizes $A_{t}$. Its particular
form will not be important for us save that for the conditions
\eqref{eq:initialcondition} it can be checked to be of the lower
triangular form. From this we can write the Wronskian-normalized basis
for wavefunctions at $z=t_0$: 
\begin{equation}
[\Phi_{t_0}(z)]_{11}= t_0^{\frac{1}{2}\theta_0}(1+{\cal
  O}(z-t_0)),\quad\quad
[\Phi_{t_0}(z)]_{12}=kt_0^{\frac{1}{2}\theta_0-1}(z-t_0)^{\theta_{t_0}}(1+{\cal O}(z-t_0)),
\end{equation}

Now, we can write the current-normalized wavefunctions in terms of the
coordinate $r$: 
\begin{equation}
\begin{gathered}
u^+_\infty=i(2\omega)^{\thalf-2iM\omega}e^{2\pi M\omega}
R^+_\infty(r)=\frac{1}{\sqrt{2\omega}}\frac{1}{r}e^{i\omega(r-r_-+2M\log(r-r_-))}
(1+{\cal O}(r^{-1})),
\\
u^-_\infty=\frac{i}{k}(2\omega)^{\thalf+2iM\omega}e^{-2\pi M\omega}
R^-_\infty(r)=\frac{1}{\sqrt{2\omega}}\frac{1}{r}e^{-i\omega(r-r_-+2M\log(r-r_-))}
(1+{\cal O}(r^{-1})),
\\
u^+_{t_0}=\frac{1}{k}e^{-i\tfrac{\pi}{2}\theta_0}\left|
\frac{2\omega t_0}{\theta_{t_0}}\right|^{\thalf}R^+_{t_0}(r)=
-i\frac{(2\omega)^{\thalf+2iM\omega}}{|t_0\theta_{t_0}|^{\thalf}}
(r-r_+)^{\thalf\theta_{t_0}}(1+{\cal O}(r-r_+)) \\
u^-_{t_0}=e^{-i\tfrac{\pi}{2}\theta_0}
\left|\frac{2\omega}{t_0\theta_{t_0}}\right|^{\thalf}
  R^-_{t_0}(r)
=\frac{(2\omega)^{\thalf(1+\theta_0)}}{|t_0\theta_{t_0}|^{\thalf}}
(r-r_+)^{-\thalf\theta_{t_0}}(1+{\cal O}(r-r_+)). 
\end{gathered}
\end{equation}
As anticipated in the previous section, a suitable choice of $k$ can
make the relative normalizations equal. The result
\eqref{eq:transmissioncoeff} follows. In order to implement the
time-reversal procedure, one has to check the relative phases and
include them explicitly in the formula for the connection
coefficients. At any rate, the basis at $r=r_+$ is
different from the one considered by \cite{Mano:1996vt,Mano:1996gn},
which were mainly concerned about generic spin Teukolsky equation and
did not work the normalization out explicitly near $r=r_+$. In the
non-zero spin case, one has to solve not only for the monodromy but
also for the Teukolsky-Starobinski identities
\cite{Chandrasekhar1983}. It is an interesting future problem to try
and extend the results presented here to that case.

\section{Angular Equation}
\label{sec:angulareq}

The general solution for the composite monodromy parameter $\sigma$
using \eqref{eq:initialdata} (and \eqref{eq:sigmastokes}) can of
course be applied to more generic situations. The associated equation
for the angular part of the wavefunction $S(\theta)$ is also of the
confluent Heun type. Following \cite{Berti:2005gp}, the scalar
spheroidal harmonics are the solutions of the angular equation: 
\begin{equation}
\frac{d}{dx}\left[(1-x^2)\frac{d}{dx}S_{\ell,m}\right]+
\left[-(a\omega)^2(1-x^2)+C_{\ell,m}+2am\omega-\frac{m^2}{1-x^2}\right]
S_{\ell,m}=0,
\label{eq:scalarang}
\end{equation}
for which the solution has the following behavior near the singular
points $x=\cos\theta=\pm 1$:
\begin{equation}
S_{\ell,m}=
\begin{cases}
(1+x)^{m/2},\quad x\rightarrow -1, \\
(1-x)^{-m/2},\quad x\rightarrow +1,
\label{eq:angularsol}
\end{cases}
\end{equation}
for integer $m$. It is clear that this behavior will only happen for
particular discrete values of the constant $C_{\ell,m}$, which can be
fed into \eqref{eq:kgAdS-4d}.

To bring this equation into the canonical confluent Heun form
\eqref{eq:heunconfluent}, we make:
\begin{equation}
z=2a\omega(x-1),\quad\quad y(z)=(1+x)^{m/2}(1-x)^{-m/2} S_{\ell,m}(x),  
\end{equation}
and the new parameters are:
\begin{equation}
t_0=-4a\omega,\quad\theta_0=-\theta_{t_0}=-m,\quad
c_0=-c_{t_0}=\frac{C_{\ell,m}+a^2\omega^2}{4a\omega}.
\end{equation}
Now, from the discussion above, the existence of a solution of the
confluent Heun equation with behavior given by \eqref{eq:angularsol}
puts constraints on the connection matrices. In this particular case,
one of the natural solutions at $z=0$ will not ``mix'' with the
natural solution at $z=1$. This would correspond to the vanishing of
one of the constants at \eqref{eq:connectiondata}. One should not that
the second natural solution will of course mix. Also, since the
parameter $\theta_0$ and $\theta_t$ are integers, there will be
logarithm behavior for the other solution -- in the case
$\omega\rightarrow 0$ it would correspond to the associated Legendre
functions of the second kind.

It is not hard to see that from this condition the composite monodromy
$2\cos\pi\sigma=\Tr M_0M_t$ will have to be special:
\begin{equation}
\sigma=2\ell, \quad\quad \ell\in\mathbb{Z},
\end{equation}
as required by the vanishing of the analogue of
\eqref{eq:transmissioncoeff} for this case. Now, using
\eqref{eq:initialdata}, one has a -- somewhat formal -- solution for
the eigenvalues:
\begin{equation}
C_{\ell,m}=-a^2\omega^2-m(m-1)+t\left. \frac{d}{dt}\log
  \tau(t;m,-m,2\ell)\right|_{t=-4a\omega}. 
\label{eq:angulareigenvalues}
\end{equation}
Since $\tau(t)$ is also a function of the Stokes parameters $s_1,s_2$, the
second equation in \eqref{eq:initialdata} is also necessary to find
the $C_\ell$.

\section{Asymptotics of Painlevé V} 
\label{sec:painleve5}

In the spirit of making the analysis self-contained, we copy the
relevant formulae about the Painlevé V $\tau$-function from
\cite{Gamayun:2013auu}. In the following we use their definition for
the $\tau$-function:
\begin{equation}
\tau(t)=t^{(\theta_0-\theta_t)^2/4}[\tilde{\tau}(t)]^{-1}, 
\end{equation}
and assume $\theta_i,\sigma$ and $\theta_i-\theta_j$,
$\sigma-\theta_i$ are not integers. The expansion for the tau-function
is of the form  
\begin{equation}
\tilde{\tau}(t,\vec{\theta})=\\
\sum_{n\in\mathbb{Z}}C(\{\theta_i\},\thalf\sigma+n)\hat{s}^nt^{(\thalf\sigma+n)^2}
{\cal B}(\{\theta_i\},\thalf\sigma+n;t),
\label{eq:tau5expansion}
\end{equation}
where the irregular conformal block ${\cal B}$ is given as a power
series over the set of Young tableaux $\mathbbold{Y}$:
\begin{equation}
{\cal
  B}(\{\theta_i\},\thalf\sigma;t)=e^{-\thalf\theta_tt}\sum_{\lambda,\mu\in\mathbbold{Y}} 
{\cal B}_{\lambda,\mu}(\{\theta_i\},\thalf\sigma)t^{|\lambda|+|\mu|}, 
\end{equation}
with coefficients
\begin{multline}
{\cal
  B}_{\lambda,\mu}=\prod_{(i,j)\in\lambda}\frac{(-\thalf\theta_\infty+\thalf\sigma+i-j)
  ((\thalf\theta_t+\thalf\sigma+i-j)^2-\tfrac{1}{4}\theta_0^2)}{h_\lambda^2(i,j)(
  \lambda'_j+\mu_i-i-j+1+\sigma)} \times \\
\prod_{(i,j)\in\mu}\frac{(-\thalf\theta_\infty-\thalf\sigma+i-j)
  ((\thalf\theta_t-\thalf\sigma+i-j)^2-\tfrac{1}{4}\theta_0^2)}{h_\mu^2(i,j)(
  \lambda_i+\mu'_j-i-j+1+\sigma)},  
\end{multline} 
where $\lambda$ denotes a Young tableau, $\lambda_i$ is the number of
boxes in row $i$, $\lambda'_j$ is the number of boxes in column $j$
and $h_\lambda(i,j)=\lambda_i+\lambda'_j-i-j+1$ is the hook length
related to the box $(i,j)\in\lambda$. The structure constants $C$
are rational products of Barnes functions
\begin{equation}
C(\{\theta_i\},\sigma)=\prod_{\epsilon=\pm}
\frac{G(1-\thalf\theta_\infty+\epsilon\thalf\sigma)
G(1+\thalf\theta_t+\thalf\theta_0+\epsilon\thalf\sigma)
G(1+\thalf\theta_t-\thalf\theta_0+\epsilon\thalf\sigma)}{ 
G(1+\epsilon\sigma)},
\end{equation}
where $G(z)$ is defined by the functional equation
$G(1+z)=\Gamma(z)G(z)$. The parameters $\sigma$ and $s$ in
\eqref{eq:tau5expansion} are related to the ``constants of
integration'' of the Painlevé V equation. In our treatment, they are
functions of the Stokes parameters $s_1$ and $s_2$. The parameter
$\sigma$ is given by \eqref{eq:sigmastokes}, whereas the expression
for $\hat{s}$ in \eqref{eq:tau5expansion} is rather long and
involved. We will outline the procedure in Section 10 of
\cite{Andreev2000} to compute it. Let $\hat{M}_0,\hat{M}_t$ be the
monodromy matrices for the hypergeometric equations. They are of the
form $\hat{M}_i=\hat{E}_i^{-1}e^{i\pi\theta_i\sigma_3}\hat{E}_i$ with
$\hat{E}_i$ given by equation (10.15) in \cite{Andreev2000}. They
satisfy $\hat{M}_t\hat{M}_0=e^{-i\pi(\theta_0+\theta_1)}{\rm
  diag}(e^{-i\pi\sigma},e^{i\pi\sigma})$. We introduce the matrix
\begin{equation}
S=
\begin{pmatrix}
-e^{-i\pi(\sigma+\theta_\infty)/2}\frac{\Gamma(-\sigma)}{
  \Gamma(1-\tfrac{1}{2}(\sigma-\theta_\infty))} & 
\frac{\Gamma(-\sigma)}{
  \Gamma(1-\tfrac{1}{2}(\sigma+\theta_\infty))} \\
e^{i\pi(\sigma-\theta_\infty)/2}\frac{\Gamma(\sigma)}{
  \Gamma(\tfrac{1}{2}(\sigma+\theta_\infty))} &
\frac{\Gamma(-\sigma)}{
  \Gamma(\tfrac{1}{2}(\sigma-\theta_\infty))}
\end{pmatrix},
\end{equation}
whose purpose is to match the diagonal monodromy of the hypergeometric
system at infinity to the non-diagonal monodromy of the confluent
hypergeometric system at the origin. It is defined up to a diagonal
matrix $s^{-\sigma_3}$. We now have the monodromy parametrization for
$E_i$ in \eqref{eq:connectionmatrices}:
\begin{equation}
E_i=\hat{E}_is^{-\sigma_3}S R^{-\sigma_3},
\end{equation}
where $R^2=s_1/s_2$ serves to bring $M_\infty$ to a symmetric
form. With this intricate construction, the parameter $s$ can be read
implicitly from $M_t=M_\infty^{-1}M_0^{-1}$ -- or more easily from
its trace. The parameter $\hat{s}$
in \eqref{eq:tau5expansion} is given by
$\hat{s}=-\tfrac{1}{2}(\sigma+\theta_\infty)s^2$. These expressions
are purely algebraic in the sense that they can be computed from the
constraints $M_\infty M_{t} M_{0}=\mathbbold{1}$ obeyed by the
monodromy matrices. The expansion \eqref{eq:tau5expansion} is obtained
from the AGT conjecture \cite{Alday:2009aq}, relating the Liouville
conformal blocks to the instanton partition function of ${\cal N}=2$
SUSY quiver theories in four dimensions via a confluence limit.

The asymptotic limit of the Painlevé V $\tau$-function was
also considered in great detail in \cite{Jimbo:1982,Andreev2000}. For
our purposes, in the limit $t_0\rightarrow 0$, the sum
\eqref{eq:tau5expansion} can be approximated by three terms:
\begin{gather}
\tilde\tau(t,\{\theta_i\},\sigma)=e^{-\thalf\theta_tt}
t^{(\sigma-2n)^2/4}f(t),\\
f(t)=K(1+B_1t+C_1t^{1-(\sigma-2n)}+{\cal O}(t^2,t^{2(1\pm(\sigma-2n))})),
\label{eq:tau5parametrization}
\end{gather}
where we assume that $2n<\sigma<2n+2$. The value of $n\in\mathbb{Z}$
will be obtained below, and we will also assume that
$\varrho=\sigma-2n<1$, which can be accomplished by shifting $n$, and
the calculation that ensues will be analogous as below. As per
\eqref{eq:tau5expansion} the leading parameters of the expansion of
$f(t)$ are:
\begin{gather}
K=C(\{\theta_i\},\thalf\sigma-n),\quad
B_1=-\frac{\theta_\infty}{4\varrho^2}(\theta_t^2-\theta_0^2+\varrho^2),\\
C_1=\frac{(\theta_\infty+\varrho)}{8\varrho^2(1-\varrho)^2} 
((\theta_t+\varrho)^2-\theta_0^2)\hat{s}^{-1}.
\end{gather}
Now, \eqref{eq:initialdata} is written as:
\begin{gather}
\tfrac{1}{4}((\theta_0-\theta_{t_0}+1)^2-\varrho^2)-t_0\frac{\dot{f}}{f}+
\tfrac{1}{2}t_0(\theta_{t_0}-1)=t_0c_{t_0}-\theta_0(\theta_{t_0}-1) \\
-t_0\frac{\dot{f}}{f}-t_0^2\frac{\ddot{f}}{f}+t_0^2\frac{\dot{f}^2}{f^2}=0.
\label{eq:initialdata2}
\end{gather}
As alluded above, the second equation does not entail information
about the initial value of the isomonodromy flow beyond the value of
$t_0$. We will see it as fixing the value of $\hat{s}$ in terms of
$\varrho$. Plugging the expansion for $f(t)$ as above, we find:
\begin{equation}
C_1=-\frac{B_1}{(1-\varrho)^2}t^{\varrho}+{\cal O}(t^{2\varrho}).
\end{equation}
Now, back to \eqref{eq:initialdata} and isolating $\varrho$, we find:
\begin{equation}
\varrho^2=(\theta_0+\theta_{t_0})^2-1+4C_{\ell,m}-2t_0+4t_0\theta_{t_0}+
4t_0\frac{\varrho}{1-\varrho}B_1+{\cal{O}}(t^{1+\epsilon}),
\end{equation}
where $\epsilon=\min(\varrho,1-\varrho)$. To first order in $t_0$, we
then have:
\begin{gather}
\varrho^2\simeq\varrho_0^2-t_0\left[2-4\theta_{t_0}-\frac{\theta_{\infty}}{
    \varrho_0(1-\varrho_0)}{((\theta_{t_0}-1)^2-\theta_0^2+\varrho_0^2)}\right], \\
\varrho_0^2=-1-16M^2\omega^2+4C_{\ell,m},
\end{gather}
where we see that the integer $n$ is chosen to offset the integer part
of $C_{\ell,m}$. Also, for small $\omega$, $\varrho_0$ is purely
imaginary, but the $t_0$ corrections will introduce a real part, and
the expansion will have poor convergence. This is verified for the
Painlevé VI expansions using formulas analogue to
\eqref{eq:tau5expansion} in \cite{Gamayun:2013auu}, as convergence of
the series improves for finite $t_0$. Notwithstanding these remarks,
we find qualitative agreement with \cite{Mano:1996vt,Castro2013b} even
though the small $t_0$ expansion is not strictly speaking a
low-frequency expansion, as other parameters also depend on $\omega$.

At any rate, the expression \eqref{eq:initialdata} should be valid for generic
values of the parameters, and with the results of Section
\ref{sec:angulareq} where the separation constant $C_{\ell,m}$ are
also computed using isomonodromy, we can claim the formal analytical
solution of the scattering problem is presented. We leave for future work a
survey of the values for the scattering coefficients, as well as
expansions for the near extremal and near-superradiance
$\theta_0\approx 0$ cases.

\section{Discussion}
\label{sec:discussion}

In this letter, we sucessfully related the scattering coefficients for
a massless scalar field in a Kerr background in terms of the Painlevé
V $\tau$-function. By choosing the current normalization of the
ingoing and outgoing waves, we first establish that the expression
\eqref{eq:transmissioncoeff} does give the transmission coefficient
directly in terms of the composite monodromy parameter $\sigma$. Also,
by relating the problem of finding $\sigma$ to the inverse
Riemann-Hilbert problem, we are able to give analytic, if implicit
solutions for the composite monodromy \eqref{eq:initialdata} and the
eigenvalue of the angular equation \eqref{eq:angulareigenvalues} in
terms of the Painlevé V $\tau$-function. Also, approximate expressions
for small isomonodromy parameter -- related to low-frequency and/or
near extremal case -- can be obtained from the asymptotics of Painlevé V
equation in terms of more elementary functions. It is our hope that the
application of the techniques outlined here and in previous work by
the authors \cite{Novaes:2014lha} will help solve many outstanding
problems in scattering for black-hole backgrounds, like Kerr-(A)dS,
and higher spin perturbations. It should be stressed that the
usefulness of \eqref{eq:initialdata} goes far beyond the low-energy or
the near horizon expansions, -- both tied to small $t_0$ -- allowing
one to study normal modes, and the onset of superradiance at
$\theta_0\approx 0$. One can also use essentially the same arguments
given here to study scalar perturbations between the inner and outer
horizon, providing yet another tool for testing for instabilities. We
will leave these important directions to future work.

The result showed here relied on the conservation of current argument,
which allowed us to equate the scattering problem to the
(mathematically defined) connection problem of the radial
equation. Although this needs proof, the same argument should work for
arbitrary spin perturbations, the so-called Teukolsky Master
Equation (TME), whose radial and angular equations for Kerr background
are also separable and yield the same confluent Heun type considered
here. This should provide a powerful analytical tool to deal with 
normal modes, generation of gravitational waves by black holes, and
the linear stability of the inner horizon.

In a more abstract viewpoint, the results extracted here point to the
striking relationship between black hole physics and Conformal Field
Theory. Relations for the $\tau$-function \eqref{eq:tau5expansion}
were obtained because of recents results concerning Liouville Field
Theories in the semi-classical limit, where the Painlevé V $\tau$-function
arises as the expansion of the irregular conformal blocks, which are
obtained from a confluence limit of (normalized) $4$-point functions in
Liouville Field Theory (see \cite{Gamayun:2013auu} and
\cite{Novaes:2014lha} for details). The $4$-point function itself
arises as the Painlevé VI $\tau$-function, and in the Black Hole case
in the Kerr-de Sitter conformally coupled scalar field scattering
\cite{daCunha:2015xxx}. The confluence limit of the Painlevé VI is the
same as the zero cosmological constant limit. This can be seen somehow
as an extension of the ideas of \cite{bredberg2010black}, where
conformal symmetry was also used to determine Black Hole scattering
in the extreme Kerr limit. As it happens, the correspondence seems
exact, at least for a wide class of (vacuum and charged) Black Holes in four
dimensions. 

While this seems surprising at first glance, one cannot help but
wonder whether this fact is yet another facet on the intricate
relationship between integrable systems and Conformal Field
Theory \cite{Hitchin2013}. General Relativity solutions like the
(static, axisymmetric) black hole systems consist an integrable sector
of Einstein equations in four dimensions
\cite{Stephani2003}. Requisition that the solutions of this integrable
sector are singularity free basically imposes that there is a
coordinate patch where the components of the metric are given in terms
of rational functions. It follows then, apart from the non-trivial
fact of separability of the equations, that the differential equations
governing the propagation of fields in these backgrounds have its
singularities easily classified, being Fuchsian or having a higher
Poincaré index. Moreover, the isomonodromy flow provides a non-linear
symmetry of these equations stemming from the theory of flat
holomorphic connections and the Painlevé property. All in all,
everything conspires to have the conformal blocks ``solving''
Fuchsian-type of differential equations, since conformal blocks are in
principle extracted from the representation theory of the Virasoro
algebra. We can then describe the chain of deep mathematical facts
pointing to the relationship between four-dimensional black holes and
Conformal Field Theory. It is a very interesting open problem to see
if this holds in the generic case.

\section*{Acknowledgements}

The authors would like to thank Marc Casals, Remo Ruffini, Monica
Guica, Amílcar de Queiroz and A. P. Balachandran for useful
discussions and comments. FN acknowledges partial support from CNPq
and BCdC from PROPESQ/UFPE. 

\appendix

\providecommand{\href}[2]{#2}\begingroup\raggedright\endgroup


\begin{thebibliography}{10}

\bibitem{Dafermos:2008en}
M.~Dafermos and I.~Rodnianski, {\it {Lectures on black holes and linear
  waves}},  {\em Clay Math.Proc.} {\bf 17} (Nov., 2013) 97--205,
  [\href{http://arxiv.org/abs/0811.0354}{{\tt arXiv:0811.0354}}].

\bibitem{Press:1972zz}
W.~H. Press and S.~A. Teukolsky, {\it {Floating Orbits, Superradiant Scattering
  and the Black-hole Bomb}},  {\em Nature} {\bf 238} (1972) 211--212.

\bibitem{York:1983zb}
J.~York, James~W., {\it {Dynamical Origin of Black Hole Radiance}},  {\em
  Phys.Rev.} {\bf D28} (1983) 2929.

\bibitem{Horowitz:1999jd}
G.~T. Horowitz and V.~E. Hubeny, {\it {Quasinormal modes of AdS black holes and
  the approach to thermal equilibrium}},  {\em Phys.Rev.} {\bf D62} (2000)
  024027, [\href{http://arxiv.org/abs/hep-th/9909056}{{\tt hep-th/9909056}}].

\bibitem{Kokkotas:1999bd}
K.~D. Kokkotas and B.~G. Schmidt, {\it {Quasinormal modes of stars and black
  holes}},  {\em Living Rev. Rel.} {\bf 2} (1999) 2,
  [\href{http://arxiv.org/abs/gr-qc/9909058}{{\tt gr-qc/9909058}}].

\bibitem{Berti:2009kk}
E.~Berti, V.~Cardoso, and A.~O. Starinets, {\it {Quasinormal modes of black
  holes and black branes}},  {\em Class.Quant.Grav.} {\bf 26} (2009) 163001,
  [\href{http://arxiv.org/abs/0905.2975}{{\tt arXiv:0905.2975}}].

\bibitem{Dolan:2008kf}
S.~R. Dolan, {\it {Scattering and Absorption of Gravitational Plane Waves by
  Rotating Black Holes}},  {\em Class. Quant. Grav.} {\bf 25} (2008) 235002,
  [\href{http://arxiv.org/abs/0801.3805}{{\tt arXiv:0801.3805}}].

\bibitem{Fiziev:2009ty}
P.~P. Fiziev and D.~R. Staicova, {\it {A new model of the Central Engine of GRB
  and the Cosmic Jets}},  {\em Bulg. Astron. J.} {\bf 11} (2009) 3--11,
  [\href{http://arxiv.org/abs/0902.2408}{{\tt arXiv:0902.2408}}].

\bibitem{Kumar:2014upa}
P.~Kumar and B.~Zhang, {\it {The physics of gamma-ray bursts \& relativistic
  jets}},  {\em Phys. Rept.} {\bf 561} (2014) 1--109,
  [\href{http://arxiv.org/abs/1410.0679}{{\tt arXiv:1410.0679}}].

\bibitem{Borissov:2009bj}
R.~S. Borissov and P.~P. Fiziev, {\it {Exact Solutions of Teukolsky Master
  Equation with Continuous Spectrum}},  {\em Bulg. J. Phys.} {\bf 37} (2010)
  065--089, [\href{http://arxiv.org/abs/0903.3617}{{\tt arXiv:0903.3617}}].

\bibitem{Leaver:1985ax}
E.~Leaver, {\it {An Analytic representation for the quasi normal modes of Kerr
  black holes}},  {\em Proc.Roy.Soc.Lond.} {\bf A402} (1985) 285--298.

\bibitem{Mano:1996vt}
S.~Mano, H.~Suzuki, and E.~Takasugi, {\it {Analytic solutions of the Teukolsky
  equation and their low frequency expansions}},  {\em Prog.Theor.Phys.} {\bf
  95} (1996) 1079--1096, [\href{http://arxiv.org/abs/gr-qc/9603020}{{\tt
  gr-qc/9603020}}].

\bibitem{Mano:1996gn}
S.~Mano and E.~Takasugi, {\it {Analytic solutions of the Teukolsky equation and
  their properties}},  {\em Prog.Theor.Phys.} {\bf 97} (1997) 213--232,
  [\href{http://arxiv.org/abs/gr-qc/9611014}{{\tt gr-qc/9611014}}].

\bibitem{compere2012kerr}
G.~Comp{\`e}re, {\it {The Kerr/CFT Correspondence and its Extensions}},  {\em
  Living Rev. Relativity} {\bf 15} (2012) 7.

\bibitem{Castro2013b}
A.~Castro, J.~M. Lapan, A.~Maloney, and M.~J. Rodriguez, {\it {Black Hole
  Scattering from Monodromy}},  {\em Class.Quant.Grav.} {\bf 30} (2013) 165005,
  [\href{http://arxiv.org/abs/1304.3781}{{\tt arXiv:1304.3781}}].

\bibitem{Novaes:2014lha}
F.~Novaes and B.~Carneiro~da Cunha, {\it {Isomonodromy, Painlevé transcendents
  and scattering off of black holes}},  {\em JHEP} {\bf 1407} (2014) 132,
  [\href{http://arxiv.org/abs/1404.5188}{{\tt arXiv:1404.5188}}].

\bibitem{Gamayun:2013auu}
O.~Gamayun, N.~Iorgov, and O.~Lisovyy, {\it {How instanton combinatorics solves
  Painlev{\'e} VI, V and IIIs}},  {\em J.Phys.} {\bf A46} (Feb., 2013) 335203,
  [\href{http://arxiv.org/abs/1302.1832}{{\tt arXiv:1302.1832}}].

\bibitem{Flammer2014}
C.~Flammer, {\em {Spheroidal wave functions}}.
\newblock Stanford University Press, Stanford, CA, 1957.

\bibitem{Berti:2005gp}
E.~Berti, V.~Cardoso, and M.~Casals, {\it {Eigenvalues and eigenfunctions of
  spin-weighted spheroidal harmonics in four and higher dimensions}},  {\em
  Phys.Rev.} {\bf D73} (2006) 024013,
  [\href{http://arxiv.org/abs/gr-qc/0511111}{{\tt gr-qc/0511111}}].

\bibitem{Neitzke:2003mz}
A.~Neitzke, {\it {Greybody factors at large imaginary frequencies}},
  \href{http://arxiv.org/abs/hep-th/0304080}{{\tt hep-th/0304080}}.

\bibitem{Andersson:2003fh}
N.~Andersson and C.~J. Howls, {\it {The Asymptotic quasinormal mode spectrum of
  nonrotating black holes}},  {\em Class. Quant. Grav.} {\bf 21} (2004)
  1623--1642, [\href{http://arxiv.org/abs/gr-qc/0307020}{{\tt gr-qc/0307020}}].

\bibitem{Motl:2003cd}
L.~Motl and A.~Neitzke, {\it {Asymptotic black hole quasinormal frequencies}},
  {\em Adv.Theor.Math.Phys.} {\bf 7} (2003) 307--330,
  [\href{http://arxiv.org/abs/hep-th/0301173}{{\tt hep-th/0301173}}].

\bibitem{Keshet:2007be}
U.~Keshet and A.~Neitzke, {\it {Asymptotic spectroscopy of rotating black
  holes}},  {\em Phys. Rev.} {\bf D78} (2008) 044006,
  [\href{http://arxiv.org/abs/0709.1532}{{\tt arXiv:0709.1532}}].

\bibitem{Jimbo1981b}
M.~Jimbo, T.~Miwa, and A.~K. Ueno, {\it {Monodromy Preserving Deformation of
  Linear Ordinary Differential Equations With Rational Coefficients, I}},  {\em
  Physica} {\bf D2} (1981) 306--352.

\bibitem{Jimbo:1981-2}
M.~Jimbo and T.~Miwa, {\it {Monodromy Preserving Deformation of Linear Ordinary
  Differential Equations with Rational Coefficients, II}},  {\em Physica} {\bf
  D2} (1981) 407--448.

\bibitem{Jimbo:1981-3}
M.~Jimbo and T.~Miwa, {\it {Monodromy Preserving Deformation of Linear Ordinary
  Differential Equations with Rational Coefficients, III}},  {\em Physica} {\bf
  D4} (1981) 26--46.

\bibitem{Jimbo:1982}
M.~Jimbo, {\it {Monodromy Problem and the boundary condition for some
  Painlev{\'e} equations}},  {\em Publ. Res. Inst. Math. Sci.} {\bf 18} (1982)
  1137--1161.

\bibitem{Alday:2009aq}
L.~F. Alday, D.~Gaiotto, and Y.~Tachikawa, {\it {Liouville Correlation
  Functions from Four-dimensional Gauge Theories}},  {\em Lett.Math.Phys.} {\bf
  91} (2010) 167--197, [\href{http://arxiv.org/abs/0906.3219}{{\tt
  arXiv:0906.3219}}].

\bibitem{Alba:2010qc}
V.~A. Alba, V.~A. Fateev, A.~V. Litvinov, and G.~M. Tarnopolskiy, {\it {On
  combinatorial expansion of the conformal blocks arising from AGT
  conjecture}},  {\em Lett.Math.Phys.} {\bf 98} (2011) 33--64,
  [\href{http://arxiv.org/abs/1012.1312}{{\tt arXiv:1012.1312}}].

\bibitem{bredberg2010black}
I.~Bredberg, T.~Hartman, W.~Song, and A.~Strominger, {\it {Black hole
  superradiance from Kerr/CFT}},  {\em Journal of High Energy Physics} {\bf
  2010} (2010), no.~4 1--32.

\bibitem{Carter:1968kx}
B.~Carter et~al., {\it {Hamilton-Jacobi and Schr{\"o}dinger separable solutions
  of Einstein's equations}},  {\em Comm. Math. Phys.} {\bf 10} (1968), no.~4
  280--310.

\bibitem{NISTSpheroidal}
H.~Volkmer, ``{Spheroidal Wave Functions}.'' {\url{http://dlmf.nist.gov/30}}.

\bibitem{Andreev2000}
F.~Andreev and A.~V. Kitaev, {\it {Connection formulas for asymptotics of the
  fifth Painlev{\'e} transcendent on the real axis}},  {\em Nonlinearity} {\bf
  13} (2000), no.~5 1801--1840.

\bibitem{Landau:1997}
L.~D.~Landau and E.~M.~Lifshitz, {\it Quantum Mechanics
  (Non-Relativistic Theory)}, {\it Course of Theoretical Physics,
  Volume 3}, 3rd Edition, Butterworth-Heinemann, 1997.

\bibitem{Iwasaki:1991}
K.~Iwasaki, H.~Kimura, S.~Shimomura, and M.~Yoshida, {\em {From Gauss to
  Painlev{\'e}: A Modern Theory of Special Functions}}, vol.~16 of {\em Aspects
  of Mathematics E}.
\newblock Braunschweig, 1991.

\bibitem{Gamayun:2012ma}
O.~Gamayun, N.~Iorgov, and O.~Lisovyy, {\it {Conformal field theory of
  Painlev\'e VI}},  {\em JHEP} {\bf 1210} (July, 2012) 038,
  [\href{http://arxiv.org/abs/1207.0787}{{\tt arXiv:1207.0787}}].

\bibitem{jimbo1980density}
M.~Jimbo, T.~Miwa, Y.~M{\^o}ri, and M.~Sato, {\it {Density matrix of an
  impenetrable Bose gas and the fifth Painlev{\'e} transcendent}},  {\em
  Physica D: Nonlinear Phenomena} {\bf 1} (1980), no.~1 80--158.

\bibitem{Segal1985}
G.~Segal and G.~Wilson, {\it {Loop groups and equations of KdV type}},  {\em
  Publications Math{\'e}matiques de l'IH{\'E}S} {\bf 61} (1985), no.~1 5--65.

\bibitem{Mason:2000aa}
L.~J. Mason, M.~A. Singer, and N.~M.~J. Woodhouse, {\it {Tau functions and the
  twistor theory of integrable systems}},  {\em Journal of Geometry and
  Physics} {\bf 32} (2000) 397--430.

\bibitem{Nekrasov:2011bc}
N.~Nekrasov, A.~Rosly, and S.~Shatashvili, {\it {Darboux coordinates, Yang-Yang
  functional, and gauge theory}},  {\em Nucl.Phys.Proc.Suppl.} {\bf 216} (Mar.,
  2011) 69--93, [\href{http://arxiv.org/abs/1103.3919}{{\tt arXiv:1103.3919}}].

\bibitem{Chandrasekhar1983}
S.~Chandrasekhar, {\it The Mathematical Theory of Black Holes},
vol. 69 in International Series of Monographs in Physics.
\newblock Oxford University Press, 1983.

\bibitem{daCunha:2015xxx}
B.~Carneiro~da Cunha and F.~Novaes, {\it {Kerr-de Sitter Greybody Factors via
  Isomonodromy}},   [\href{http://arxiv.org/abs/1508.04046}{{\tt
  arXiv:1508.04046}}].

\bibitem{Hitchin2013}
N.~J. Hitchin, G.~B. Segal, and R.~S. Ward, {\em {Integrable systems: Twistors,
  loop groups, and Riemann surfaces}}.
\newblock Oxford University Press, 2013.

\bibitem{Stephani2003}
H.~Stephani, D.~Kramer, M.~MacCallum, C.~Hoenselaers, and E.~Herlt, {\em {Exact
  solutions of Einstein's field equations}}.
\newblock Cambridge University Press, 2003.

\end{thebibliography}
\end{document}